# First-principles investigation of half-metallic ferromagnetism of half-Heusler compounds XYZ

L. Feng[a,*]  E. K. Liu[b], W. X. Zhang[a], W. H. Wang[b], G. H. Wu[b]

[a]*Key Laboratory of Advanced Transducers and Intelligent Control System, Ministry of Education, Computational Condensed Matter Physics Laboratory, Department of Physics, Taiyuan University of Technology, Taiyuan 030024, People's Republic of China*

[b]*Beijing National Laboratory for Condensed Matter Physics, Institute of Physics, Chinese Academy of Sciences, Beijing 100190, People's Republic of China*

**Abstract:** We investigate the electronic structure and magnetism of half-Heusler compounds XYZ (X, Y=V, Cr, Mn, Fe, Co and Ni; Z=Al, Ga, In, Si, Ge, Sn, P, As, and Sb) using the ab initio density functional theory calculations. Nine half-metals with half-Heusler structure have been predicted with the half-metallic gap of 0.07-0.67 eV. The calculations show that the formation energies for these nine half-Heusler compounds range from -1.32 to -0.12 eV/f.u., and for CoCrSi, CoCrGe, CoFeGe, CoMnSi, CoMnGe, FeMnGe and FeMnAs, the total energy differences between the half-Heusler structure and the corresponding ground-state structure are small (0.07-0.76 eV/f.u.), thus it is expected that they would be realized in the form of thin films under metastable conditions for spintronic applications. The stability of the half-metallicity of CoCrGe and FeMnAs to the lattice distortion is also investigated in detail.

## 1. Introduction

Spintronics is an emerging field in nanoscale electronics which uses the spin of electrons, rather than an electric charge, to encode and process data [1-3]. Half-metallic materials with complete spin polarization at the Fermi level are highly attractive for spintronics applications because of their high spin polarization. This kind of materials has great attraction to scientific researchers due to their potential device applications such as nonvolatile magnetic random access memories (MRAM) and magnetic sensors [4, 5]. The concept of half-metallic ferromagnets was first introduced by de Groot et al. [6, 7], on the basis of band structure calculations in NiMnSb and PtMnSb half-Heusler compounds. Half-metallic materials have been found theoretically in many materials, for example ferromagnetic metallic oxides [8-10], dilute magnetic semiconductors [11, 12], zincblende compounds [13-17], full-Heusler compounds [18-23] and half-Heusler compounds [24-26].

---
[*] Corresponding author.
E-mail address:fenglin@tyut.edu.cn

Half-Heusler compounds, which have the chemical formula XYZ, crystallize in the face-centered cubic $C1_b$ structure with the space group F-43m [24]. In this structure, X, Y and Z atoms occupy (1/4,1/4,1/4), (0,0,0) and (1/2,1/2,1/2) sites, respectively, and (3/4,3/4,3/4) site is empty. Considering half-metal with half-Heusler structure exhibited high Curie temperature on the one hand, and on the other hand, their lattices match well with many semiconductor substrates such as MgO and GaAs, thus it is meaningful to investigate the structural stability and potential half-metallicity of half-Heusler compounds XYZ (X, Y=V, Cr, Mn, Fe, Co and Ni; Z=Al, Ga, In, Si, Ge, Sn, P, As, and Sb). In this paper, in order to explore the potential half-metallicity of these compounds, we use the first-principles calculations to systematically investigate their electronic, magnetic and structural stability. Nine half-Heusler half-metals are predicted with the half-metallic gap of 0.07-0.67 eV. In fact, most of these nine compounds crystallize in hexagonal $Ni_2In$ structure in the bulk form [27-30], and some of them exhibited the property of phase transformation, forming orthorhombic TiNiSi structure [31-35]. In order to investigate the possibility of experimentally fabrication of these half-Heusler half metals, we further reveal from the structural stability that the formation energies for these nine half-Heusler compounds are negative, thus suggesting a possibility of synthesizing these half-metals in the experiment. For CoCrSi, CoCrGe, CoFeGe, CoMnSi, CoMnGe,

FeMnGe and FeMnAs, the total energy differences between the half-Heusler structure and the corresponding ground-state structure are small (0.07-0.76 eV/f.u.), thus it is expected that they would be realized in the form of thin films under metastable conditions for spintronic applications.

## 2. Computational details

The spin-polarized density functional theory (DFT) calculations are performed with Cambridge Serial Total Energy (CASTEP) code [36,37]. The interaction between ions and electrons is described by ultrasoft pseudopotentials [38,39]. The generalized gradient approximation (GGA) in Perdew and Wang parameterization [40-42] is used to describe the exchange correlation energy. For the pseudopotentials used, the electronic configurations with core level correction are $Cr(3d^4 4s^2)$, $Mn(3d^5 4s^2)$, $Fe(3d^6 4s^2)$, $Co(3d^7 4s^2)$, $Si(3s^2 3p^1)$, $Ge(4s^2 4p^2)$, $P(3s^2 3p^3)$ and $As(4s^2 4p^3)$, respectively. The cut-off energy of the plane wave basis set was 400 eV for all of the cases, and 120, 144, 112 and 112 k-points were employed in the irreducible Brillouin zone of half-Heusler structure, hexagonal $Ni_2In$ structure, orthorhombic TiNiSi structure and tetragonal $Fe_2As$ structure, respectively. These parameters ensure good convergences for the total energy. The convergence tolerance for the calculations was selected as the difference in the total energy within the $10^{-6}$ eV/atom.

## 3. Results and discussion

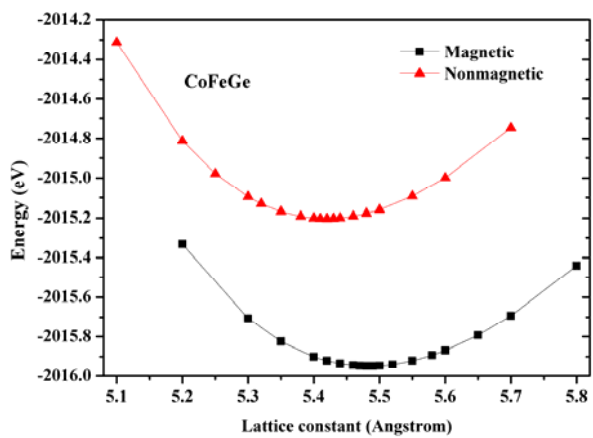

Figure 1(color online) Calculated total energy of CoFeGe with half-Heusler structure as a function of the lattice constants for magnetic and nonmagnetic states

In the first step, the lattice constants were determined. The total energies for XYZ compounds with half-Heusler structure as a function of the lattice constant for both nonmagnetic state and magnetic states are calculated. The magnetic state is found to be more stable in energy than the nonmagnetic one. The equilibrium lattice constants were derived by minimizing the total energy. As an example, the lattice constant dependence of the total energy for CoFeGe is shown in Fig. 1. The predicted equilibrium lattice constants and the energy differences between the magnetic and nonmagnetic states are summarized in Table 1. We further calculate the total density of states (DOS) in the magnetic state with the equilibrium lattice constants. Fig. 2 and Fig. 3(b) presents the spin-polarized DOSs of the nine compounds at their equilibrium lattice constants. For these nine compounds, the electronic states in the majority-spin band are metallic, and there is an energy gap at the Fermi level for the minority-spin band. So they are half-metals. Furthermore, the calculated total magnetic moment per formula unit (see Table 1) is integral, which is a typical characteristic of half-metallic (HM) ferromagnets [43].

**Table 1** Equilibrium lattice parameter ($a_0$), energy difference between magnetic state and nonmagnetic state ($E_{M-N}$), total magnetic moment ($m^t$), atom-resolved magnetic moment ($m^X$ and $m^Y$), band gap energy ($E_{bg}$) and half-metallic gap ($E_{HMg}$) for the nine compounds with half-Heusler structure.

| compounds | $a_0$(Å) | $E_{M-N}$(eV) | $m^t(\mu_B)$ | $m^X(\mu_B)$ | $m^Y(\mu_B)$ | $E_{bg}$(eV) | $E_{HMg}$(eV) |
|---|---|---|---|---|---|---|---|
| CoCrSi | 5.39 | -0.14 | 1.00 | -0.42 | 1.56 | 0.67 | 0.66 |
| CoCrGe | 5.49 | -0.19 | 1.00 | -0.58 | 1.72 | 0.81 | 0.55 |
| CoFeGe | 5.49 | -0.73 | 3.00 | 0.58 | 2.48 | 0.18 | 0.16 |
| CoMnSi | 5.41 | -0.44 | 2.00 | -0.04 | 2.12 | 0.64 | 0.19 |
| CoMnGe | 5.53 | -0.53 | 2.00 | -0.24 | 2.34 | 0.80 | 0.07 |
| FeMnGe | 5.47 | -0.12 | 1.00 | -0.72 | 1.74 | 0.39 | 0.25 |
| FeMnP | 5.32 | -0.27 | 2.00 | -0.26 | 2.24 | 0.68 | 0.53 |
| FeMnAs | 5.50 | -0.42 | 2.00 | -0.56 | 2.54 | 0.90 | 0.42 |
| FeCrAs | 5.48 | -0.13 | 1.00 | -0.60 | 1.80 | 0.88 | 0.67 |

**A. half-metallicity in half-Heusler phase**

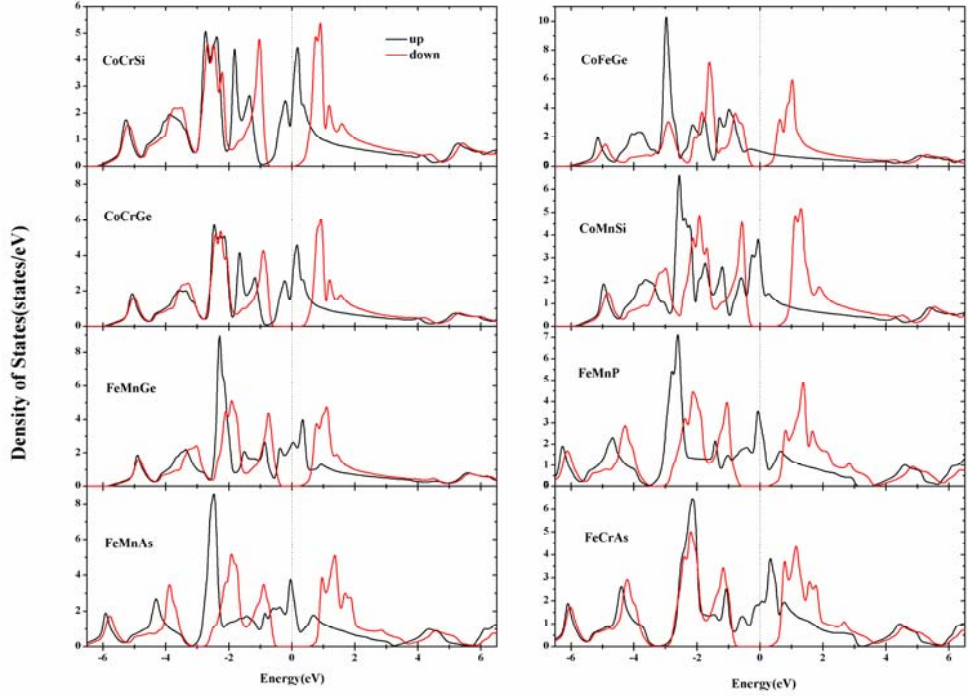

Figure 2(color online) Spin-polarized total density of states for CoCrSi, CoCrGe, FeMnGe, FeMnAs, CoFeGe, CoMnSi，FeMnP and FeCrAs at their equilibrium lattice constants

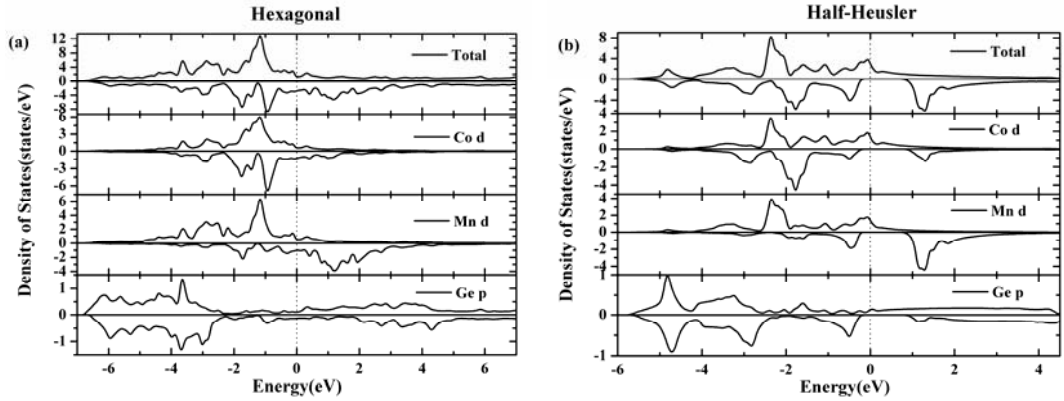

Figure 3(color online) Calculated spin-projected total and partial density of states of CoMnGe in (a) hexagonal $Ni_2In$ structure and (b) half-Heusler structure

It has been mentioned that most of the compounds listed in Table 1 crystallize in hexagonal $Ni_2In$ structure in the bulk form, and some of them exhibit the property of phase transformation, forming orthorhombic TiNiSi structure. Therefore, the nine compounds do not exhibit half-metallicity in the experiments up to now. However, according to our calculations the half-metallicity appears when they crystallize in half-Heusler structure. Here, we take CoMnGe as an example to make a comparison. The calculated total DOSs and PDOSs of CoMnGe for hexagonal $Ni_2In$ structure and half-Heusler structure are presented in Fig. 3. It is obvious that for hexagonal $Ni_2In$ structure both the majority-spin and minority-spin bands are metallic. The calculated electronic structure for hexagonal $Ni_2In$ structure CoMnGe is consistent with the previous study by Kaprzyk et al. [44]. For hexagonal $Ni_2In$ structure, the majority- and minority-spin bands, which range from −6 eV to −2 eV, are due to the *p-d* hybridization of Ge *4p* states with Co *3d* and Mn *3d* states. The upper majority- and minority spin dispersed bands are mainly composed of hybridized Co *3d* and Mn *3d* states. However, this *d-d* hybridization does not give rise to the energy gap in the Fermi level. For the half-Heusler structure, in the majority-spin component, Mn *3d* states are mostly occupied and hybridized with Co *3d* electrons; in the minority spin part, local and mostly nonhybridized Mn *3d* states are found at about 1 eV above Fermi level. The origin of the gap is mainly attributed to the covalent hybridization between the *d* states of the Co and Mn atoms, leading to the formation of bonding and antibonding bands with a gap in between [45]. The bonding hybrids are localized mainly at the Co atoms whereas the antibonding states are mainly at the Mn sites. The half-metallic gaps ($E_{HMg}$), the most proper indication of the half-metallicity of a material, are also presented in Table 1. Here, the half-metallic gap is determined as the minimum between the bottom energy of minority (majority) spin conduction bands with respect to the Fermi level and the absolute values of the top energy of minority (majority) spin valence bands. The $E_{HMg}$s of CoCrSi, CoCrGe, FeMnP, FeMnAs and FeCrAs are 0.66, 0.55, 0.53, 0.42 and 0.67 eV, respectively. These values are much larger than that of NiMnSb, FeMnSb [46] and CoMnSb [47], thus they are more stable in practical applications than the ones with a small $E_{HMg}$.

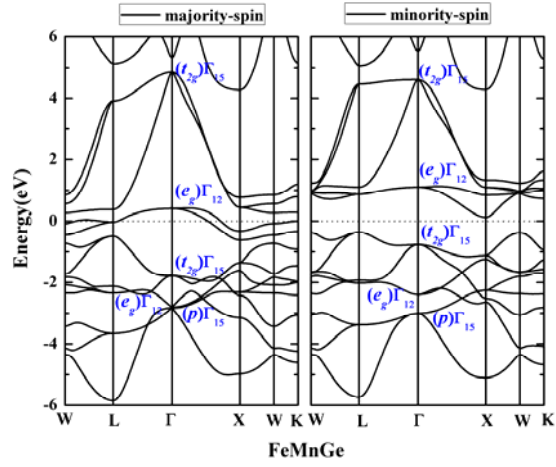

Figure 4(color online) Spin-polarized band structure of FeMnGe with half-Heusler structure

In order to investigate the electronic structure in detail, we further show the spin-polarized band structure for FeMnGe in Fig. 4. In the majority-spin bands, at about -3.0 eV the two fully-filled $\Gamma_{12}$ bands mainly originate from the hybridized $d$-$e_g$ electrons of Fe and Mn (Here, the three $\Gamma_{15}$ bands originating from $p$ electrons of Ge overlap with them). The higher three fully-filled $\Gamma_{15}$ bands at about -2.0 eV mainly originate from the hybridized $d$-$t_{2g}$ electrons of Fe and Mn. The two narrow $\Gamma_{12}$ bands crossing the Fermi level are mainly occupied by the $d$-$e_g$ electrons of Fe and Mn. On the other hand, in the minority-spin bands the three $\Gamma_{15}$ bands just below $E_F$ are created by the bonding $t_{2g}$ states of Fe and Mn with a small part of Ge $p$ states. Above the Fermi level the two narrow antibonding $\Gamma_{12}$ bands and three antibonding $\Gamma_{15}$ bands can be found. The covalent hybridization between the 3$d$ states of the Fe and Mn atoms pushes the minority-spin $\Gamma_{12}$ bands above the Fermi level, opening the minority-spin energy gap at the Fermi level, and meanwhile pulls the majority-spin $\Gamma_{12}$ bands cross the Fermi level, showing metallicity in the majority-spin bands, which results in the HM ferromagnetism in FeMnGe. In the minority-spin band structure of FeMnGe, the valence band maximum (VBM) is at the W-point and the conduction band minimum (CBM) at the X-point. Thus, the minority-spin band structure shows semiconducting behavior with an indirect energy gap. The band gap energies for these nine compounds are summarized in Table 1.

Here, we come to the magnetic properties of these half-Heusler compounds. As Table 1 shows, the magnetic moment calculations show that total magnetic moment per formula unit for all of these nine compounds are integral. The integral total magnetic moment, which is a typical characteristic of HM ferromagnets, obeys the Slater-Pauling rule for the half-Heusler alloys [48]: $M_t = Z_t - 18$; where $M_t$ is the total magnetic moment (in $\mu_B$) per formula unit and $Z_t$ is the total number of valence electrons. For example, Fe, Cr and As atoms have 8, 6 and 5 valence electrons, respectively. The total number of valence electrons $Z_t$ is 19, thus the total magnetic moment per formula unit is 1 $\mu_B$.

**B. Structural stability, Potential to be realized and Stability of half-metallicity**

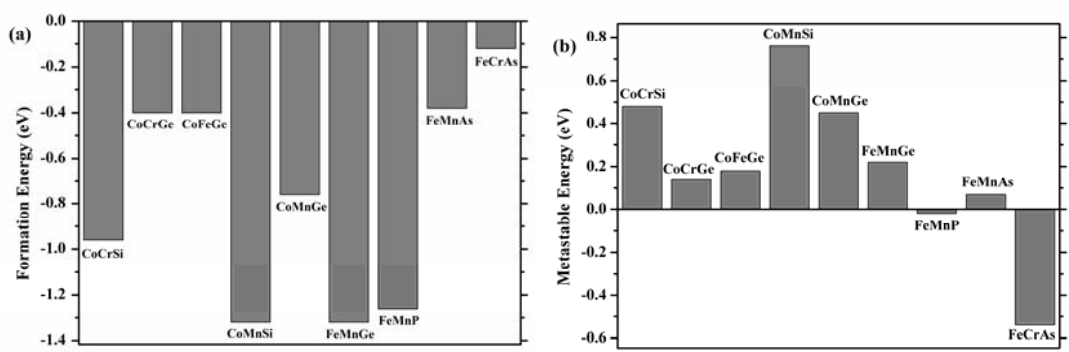

Figure 5 (a) Formation energies for the nine half-Heusler compounds (b) Metastable energies for the nine half-Heusler compounds

In this section, we further explore the structural stability of these nine HM compounds by calculating their formation energies and metastable energies. The formation energy is defined as the total energy per formula unit with respect to the sum of the elemental energies. The calculated formation energies for these nine compounds are presented in Fig. 5(a). We can find that the formation energies for all these nine compounds are negative and the absolute values are considerable, thus suggesting a possibility of synthesizing these half-metals in the experiment.

The metastable energy is defined as the energy difference between the half-Heusler structure and the corresponding ground state structure. The metastable energies for the nine compounds with the half-Heusler structure are illustrated in Fig. 5(b). For CoCrGe, CoFeGe, FeMnGe, FeMnP and FeCrAs, the ground state is hexagonal $Ni_2In$ structure [27-30]. Their metastable energies for the half-Heusler structure are 0.14, 0.18, 0.22, -0.02 and -0.54 eV, respectively. Here, the calculated ground state for FeMnP and FeCrAs is half-Heusler structure, which is not consistent with the experiments. That might be due to the fact that in intermetallic compounds the most stable structure is not necessarily the structure with the lowest energy. The other factors such as bond length and electronegativity also have an effect on the stability of the structure. For CoMnSi and CoMnGe, the ground state is orthorhombic TiNiSi structure [31,32]. Their metastable energies for the half-Heusler structure are 0.76 and 0.45 eV, respectively. For FeMnAs, the ground state is tetragonal $Fe_2As$ structure [49]. Its metastable energy for half-Heusler structure is 0.07 eV. There is no experimental report about CoCrSi. According to our calculations, its ground state is orthorhombic TiNiSi structure, and its metastable energy for the half-Heusler structure is 0.48 eV.

The zincblende pnictides, MnAs, CrAs and CrSb are predicted to be half-metals [50]. Their ground state is hexagonal NiAs structure. Though the matastable energies for zincblende MnAs, CrAs and CrSb are 0.90, 0.90 and 1.00 eV [51], respectively, they could be successfully fabricated in the form of thin films in experiments. However, the best thin film samples, achieved in the case of the CrAs, are limited to only 5 unit cells in thickness [52]. On the other hand, the research work by Zhao et al. has indicated that when the energy difference between the zincblende and NiAs structure in the bulk form is less than 0.30 eV, the ground state for epitaxial film could be the zincblende structure [53]. In this case, the thickness of the film could break through the limitation of 5 unit cells. Since the zincblende and NiAs structure are similar to the half-Heusler and $Ni_2In$ structure, it is reasonable to investigate the structure stability of XYZ compounds with half-Heusler in the same way as the zincblende pnictides. For CoCrSi, CoMnSi and CoMnGe with half-Heusler structure, their metastable energies are greater than 0.30 eV. They might be fabricated in the form of thin films, but the thickness will be very small. However, the metastable energies for CoCrGe, CoFeGe, FeMnGe and FeMnAs with half-Heusler structure are all less than 0.30 eV, thus they could be fabricated in the form of thin films under appropriate substrate, and the thickness will be considerable.

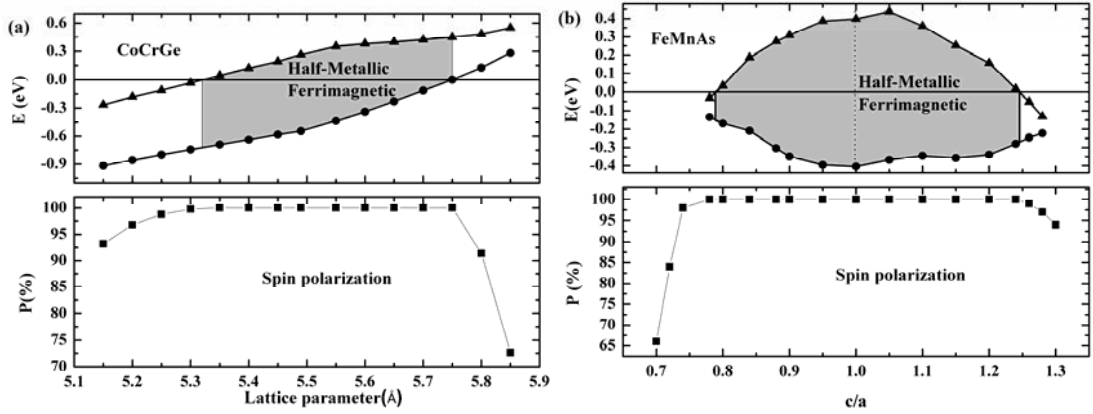

Figure 6(color online) (a) The half-metallic state and spin polarization as a function of the lattice constant for CoCrGe with half-Heusler structure. (b) The half-metallic state and spin polarization as a function of the c/a ratio for FeMnAs with half-Heusler structure

In practical applications, the lattice of the half-metallic materials is usually distorted, which may affect their half-metallic character. In order to investigate the stability of the half-metallicity to the lattice constant change in detail, we calculated the electronic structures of these half-Heusler compounds for lattice distortion. Fig. 6(a) illustrates the half-metallic state and the spin polarization as a function of the lattice constant for CoCrGe in the range of 5.15–5.85 Å. The half-metallic region for CoCrGe is in the range of 5.32–5.75 Å. That means CoCrGe can maintain its half-metallicity when its lattice constants are changed by -3.1%–4.7% relative to the equilibrium lattice constant. The half-metallicity of CoCrGe are more robust against lattice constant change than that of NiCrP (-3%–2%), NiCrTe (-1.5%–3%), NiCrSe (0%–3.5%) [54] and NiMnSb (-2%–3%) [55].

In the growth of thin films the tetragonal distortion with fixed volume is most likely to occur. Therefore, it is meaningful to investigate the relationship between the spin polarization and the tetragonal distortion for these compounds with half-Heusler structure. Fig. 6(b) illustrates the half-metallic state and the spin polarization as a function of the c/a ratio for FeMnAs in the range of 0.70–1.30. It appears that the half-metallic character exhibits a low sensitivity to the tetragonal distortion in the lattice variation range of ±20% around the equilibrium lattice parameter for FeMnAs. In particular, FeMnAs could maintain its half-metallicity when its lattice parameters are a=b=5.69 Å, c=5.12 Å (c/a=0.9). This in-plane parameter is very close to the lattice parameter of GaAs, and its formation energy are considerable while the metastable energy is very small, thus FeMnAs with half-Heusler structure is very promising to be fabricated in the form of thin films on GaAs substrate. Here, we just choose half-Heusler CoCrGe and FeMnAs to illustrate the stability of the half-metallicity and the half-metallicities of the other seven compounds are also robust against the lattice constant change.

## 4. Conclusions

We investigate the electronic structure and magnetism of half-Heusler compounds XYZ (X, Y=V, Cr, Mn, Fe, Co and Ni; Z=Al, Ga, In, Si, Ge, Sn, P, As, and Sb) using the ab initio density functional theory calculations. Nine half-Heusler compounds are predicted to be HM ferromagnets with the half-metallic gap of 0.07-0.67 eV. The calculations show that the formation energies for these compounds with half-Heusler structure range from -1.32 to -0.12 eV/f.u., and for CoCrSi, CoCrGe, CoFeGe, CoMnSi, CoMnGe, FeMnGe and FeMnAs, the total energy differences between the half-Heusler structure and the corresponding ground-state structure are small (0.07-0.76 eV/f.u.), thus it is expected that they would be realized in the form of thin films under metastable conditions for spintronic applications. Their half-metallicities are robust against lattice distortion and FeMnAs is suitable to be fabricated in the form of thin films on GaAs substrate.

**Acknowledgement:**

This work is supported by the National Natural Science Foundation of China in Grant Nos. 51301119, 51301195 and 11204201, and the Natural Science Foundation for Young Scientists of Shanxi in Grant No. 2013021010-1.

Captions:

Figure 1(color online) Calculated total energy of CoFeGe with half-Heusler structure as a function of the lattice constants for magnetic and nonmagnetic states

Figure 2(color online) Spin-polarized total density of states for CoCrSi, CoCrGe, FeMnGe, FeMnAs, CoFeGe, CoMnSi，FeMnP and FeCrAs at their equilibrium lattice constants

Figure 3(color online) Calculated spin-projected total and partial density of states of CoMnGe in (a) hexagonal $Ni_2In$ structure and (b) half-Heusler structure

Figure 4(color online) Spin-polarized band structure of FeMnGe with half-Heusler structure

Figure 5 (a) Formation energies for the nine half-Heusler compounds (b) Metastable energies for the nine half-Heusler compounds

Figure 6(color online) (a) The half-metallic state and spin polarization as a function of the lattice constant for CoCrGe with half-Heusler structure. (b) The half-metallic state and spin polarization as a function of the c/a ratio for FeMnAs with half-Heusler structure